\pdfoutput=1
\documentclass[sigconf]{acmart}
\acmConference[ICSE 2024]{46th International Conference on Software Engineering}{April 2024}{Lisbon, Portugal}
\AtBeginDocument{%
  \providecommand\BibTeX{{%
    \normalfont B\kern-0.5em{\scshape i\kern-0.25em b}\kern-0.8em\TeX}}}

\usepackage{framed}
\usepackage{listings}
\usepackage{xcolor}
\usepackage{color}
\usepackage{multirow}
\usepackage{algorithm}

\usepackage{amsmath}
\usepackage{amssymb}
\usepackage{threeparttable}
\usepackage{url}
\usepackage[most]{tcolorbox}
\usepackage{hyperref}
\usepackage[hyphenbreaks]{breakurl}
\usepackage{algpseudocode}
\usepackage{colortbl}
\hypersetup{
    colorlinks = true,
    linkcolor=blue,
    filecolor=blue,      
    urlcolor=blue,
    citecolor=cyan,
}
\usepackage{xspace}
\usepackage{subfigure}
\usepackage{graphicx}
\usepackage{fvextra}

\usepackage[shortlabels]{enumitem}
\setlist[enumerate]{nosep}

\definecolor{codegreen}{rgb}{0,0.6,0}
\definecolor{codegray}{rgb}{0.5,0.5,0.5}
\definecolor{codepurple}{rgb}{0.58,0,0.82}
\definecolor{backcolour}{rgb}{0.95,0.95,0.92}
\definecolor{lightgreen}{rgb}{0,0.4,0}
\definecolor{lightred}{rgb}{0.4,0,0}

\definecolor{mygray}{gray}{.9}

\lstdefinestyle{mystyle}{
    commentstyle=\color{brown},
    keywordstyle=\color{magenta},
    numberstyle=\tiny\color{codegray},
    stringstyle=\color{codepurple},
    basicstyle=\ttfamily\footnotesize,
    breakatwhitespace=false,         
    breaklines=true,                 
    captionpos=b,                    
    keepspaces=true,                 
    numbers=left,                    
    numbersep=5pt,                  
    showspaces=false,                
    showstringspaces=false,
    showtabs=false,                  
    tabsize=2,
    escapeinside={<@}{@>},
    language=python
}

\newcommand{\tool}{\textsc{TypeFix}\xspace}
\newcommand{\typebugs}{\textsc{TypeBugs}\xspace}
\newcommand{\bugsinpy}{\textsc{BugsInPy}\xspace}

\newcommand{\definition}[2]{\textbf{Definition #1 (#2).}}

\newcommand\etal{{\it{et al.\ }}}

\newcommand{\tabincell}[2]{\begin{tabular}{@{}#1@{}}#2\end{tabular}}
\newcommand{\eat}[1]{\if 0 #1 \fi}

\newcommand{\g}{\cellcolor{mygray}}

\newfont{\mycrnotice}{ptmr8t at 7pt}
\newfont{\myconfname}{ptmri8t at 7pt}

\newcommand{\answer}[2]{
\begin{tcolorbox}[breakable,width=\linewidth,boxrule=0pt,top=1pt, bottom=1pt, left=1pt,right=1pt, colback=gray!20,colframe=gray!20]
\textbf{Answer to RQ#1:} #2
\end{tcolorbox}
}

\begin{document}

\title{Domain Knowledge Matters: Improving Prompts with Fix Templates for Repairing Python Type Errors }

\author{Yun Peng}
 \affiliation{\institution{The Chinese University of Hong Kong}\country{Hong Kong, China}}
 \email{ypeng@cse.cuhk.edu.hk}

\author{Shuzheng Gao}
 \affiliation{\institution{The Chinese University of Hong Kong}\country{Hong Kong, China}}
 \email{1155203205@link.cuhk.edu.hk}

\author{Cuiyun Gao}
 \affiliation{\institution{Harbin Institute of Technology} \country{Shenzhen, China}}
 \email{gaocuiyun@hit.edu.cn}
 \authornote{Corresponding author}

\author{Yintong Huo}
 \affiliation{\institution{The Chinese University of Hong Kong}\country{Hong Kong, China}}
 \email{ythuo@cse.cuhk.edu.hk}

\author{Michael R. Lyu}
 \affiliation{\institution{The Chinese University of Hong Kong} \country{Hong Kong, China}}
 \email{lyu@cse.cuhk.edu.hk}

\begin{abstract}
As a dynamic programming language, Python has become increasingly popular in recent years. Although the dynamic type system of Python facilitates the developers in writing Python programs, it also brings type errors at run-time which are prevalent yet not easy to fix. There exist rule-based approaches for automatically repairing Python type errors. The approaches can generate accurate patches for the type errors covered by manually defined templates, but they require domain experts to design patch synthesis rules and suffer from low template coverage of real-world type errors. Learning-based approaches alleviate the manual efforts in designing patch synthesis rules and have become prevalent due to the recent advances in deep learning. Among the learning-based approaches, the prompt-based approach which leverages the knowledge base of code pre-trained models via pre-defined prompts, obtains state-of-the-art performance in general program repair tasks. However, such prompts are manually defined and do not involve any specific clues for repairing Python type errors, resulting in limited effectiveness.
How to automatically improve prompts with the domain knowledge for type error repair is challenging yet under-explored.

In this paper, we present \tool, a novel prompt-based approach with fix templates incorporated for repairing Python type errors. \tool first mines generalized fix templates via a novel hierarchical clustering algorithm. The identified fix templates indicate the common edit patterns and contexts of existing type error fixes. \tool then generates code prompts for code pre-trained models by employing the generalized fix templates as domain knowledge, in which the masks are adaptively located for each type error instead of being pre-determined. Experiments on two benchmarks, including \textsc{BugsInPy} and \textsc{TypeBugs}, show that \tool successfully repairs 26 and 55 type errors, outperforming the best baseline approach by 9 and 14, respectively. Besides, the proposed fix template mining approach can cover 75\% of developers' patches in both benchmarks, increasing the best rule-based approach PyTER by more than 30\%.
\end{abstract}


%\keywords{To be filled.}


\maketitle

\section{Introduction}\label{sec:intro}

Being used in most artificial intelligence and data science applications, Python becomes extremely popular in recent years. According to GitHub Octoverse~\cite{octoverse}, which records the state of open-source software, Python is the second most-used programming language in 2022. Moreover, Python continues to see gains in its usage across GitHub with a 22.5\% year-over-year increase~\cite{octoverse}.

Python adopts a dynamic type system, in which the type of a variable will be resolved only at run-time. This enables fast prototyping and brings much convenience for developers to write an executable program. The catch, however, is that more type errors occur at run-time, threatening the reliability of Python applications. Oh \etal~\cite{pyter} find that about 30\% of questions in Stack Overflow and issues in GitHub of Python are about type errors. To avoid type errors, Python Software Foundation accepts several Python Enhancement Proposals (PEPs)~\cite{pep484, pep544, pep585, pep589} and releases a static type checker named \textit{mypy}~\cite{mypy}, allowing developers to add type annotations and check potential type conflicts statically. What's more, the recent research~\cite{allamanis20typilus, peng22static, mir22type4py, sun22static} on type inference aims at statically inferring the types of variables, which further reduces the burden of manual type annotation. These approaches can reduce potential type errors but provide limited help to repair existing type errors.

To automatically fix type errors, Oh \etal~\cite{pyter} propose the first rule-based approach. They manually define nine templates and several synthesis rules to generate patches via dynamic analysis, but the manually defined templates suffer from low coverage of real-world type errors and designing patch synthesis rules requires substantial efforts from domain experts. 

General learning-based automatic program repair (APR) approaches~\cite{Lutellier20coconut,jiang21cure,drain21deepdebug,zhu21a,xia22less,ye22neural} become quite popular and powerful in recent years, since they are feature-agnostic and can automatically learn to generate patches from existing bug fixes, without explicit definitions of synthesis rules. Among the learning-based approaches, the Neural Machine Translation (NMT)-based approach that translates the buggy lines into correct lines was typically used in the past. Most recently, Xia \etal~\cite{xia22less} propose the first prompt-based APR approach named \textit{AlphaRepair} and obtain state-of-the-art performance. Unlike NMT-based APR approaches, AlphaRepair transforms the APR problem into a fill-in-the-blank problem by masking several tokens in buggy lines and invoking pre-trained models to predict the masked tokens.
Despite the superior performance of the prompt-based approach over NMT-based approaches, the prompts in AlphaRepair are pre-defined, i.e., where to mask and how to add masks in buggy code are manually designed. Without domain-specific knowledge, the prompt-based approach can hardly fix the type errors with complex patterns~\cite{pyter}. However, automatically incorporating the prompt with domain knowledge is challenging due to different levels of type errors and various type error fixing patterns.

\textbf{Our Work.} To address the aforementioned challenge, we propose \tool, a domain-aware prompt-based approach for repairing type errors. \tool has two main phases: the template mining phase and the patch generation phase. The template mining phase aims at extracting and organizing fix templates from existing type error fixes. Fix templates are designed to handle type errors at different levels (e.g., expression level and statement level). \tool first parses type error fixes into \textit{specific fix templates} and then employs a novel hierarchical clustering algorithm to abstract and merge the \textit{specific fix templates} into \textit{general fix templates}. The patch generation phase aims at exploiting the mined fix templates in the first phase for producing patches. \tool selects the most matched and commonly-used fix templates based on Breadth-First Search (BFS) and a frequency-aware ranking algorithm, and then generates code prompts by applying the ranked fix templates, and invokes CodeT5~\cite{wang21codet5} for prediction. \tool is fully automated and extendable, as it does not need manually defined templates as well as patch synthesis rules. Additionally, the minded fix templates enable the proposed prompt-based \tool to be aware of domain knowledge when generating patches.

We evaluate \tool on two benchmarks \textsc{BugsInPy}~\cite{bugsinpy} and \textsc{TypeBugs}~\cite{pyter} by comparing it with four baselines including both the recent rule-based and learning-based approaches. In the \textsc{BugsInPy} benchmark, \tool successfully fixes 26 out of 54 type errors, outperforming the most effective baseline Codex~\cite{codex} by 9. In the \textsc{TypeBugs} benchmark, \tool successfully fixes 55 out of 109 bugs, outperforming the most effective baseline PyTER~\cite{pyter} by 14. Experiments also show that the fix templates mined by \tool can cover about 75\% of type errors in both benchmarks, much higher than PyTER which only covers 40\% of the type errors. The results demonstrate the effectiveness of \tool in repairing Python type errors.

\textbf{Contributions.} We conclude our contributions as follows.
\begin{itemize}
    \item To the best of our knowledge, \tool is the first domain-aware prompt-based approach for repairing Python type errors.
    \item We propose a novel fix template design that can handle type errors at different levels, along with a novel hierarchical clustering approach to mine various fix templates from existing type error fixes.
    \item Extensive experiments demonstrate the effectiveness of \tool compared with state-of-the-art rule-based and learning-based baselines, and the high coverage of the mined fix templates.
\end{itemize}

\section{Motivation}~\label{sec:mot}

To better illustrate our motivation, we give an example in Listing~\ref{lst:motex}. The type error in Listing~\ref{lst:motex} is from a popular GitHub project \textit{scrapy} in the \textsc{BugsInPy} benchmark. The correct fix for this type error is to add a user-defined type conversion function \textit{to\_bytes} to the entire string, as shown in the green-colored line. We also provide the patches provided by the baseline approaches and \tool in Listing~\ref{lst:motex}.

\textbf{Baseline Approaches.} CoCoNuT~\cite{Lutellier20coconut} is an NMT-based APR approach that translates the buggy line into the correct line. In the patch, it modifies the content of the string since the variable \textit{user} and \textit{password} are often used in authorization. However, this cannot fix the type error. AlphaRepair~\cite{xia22less} is a prompt-based APR approach that masks the tokens in the buggy line to generate patches. In the patch, it masks the function name \textit{unquote} and fills a new name \textit{ascii} to generate the patch. Without the domain knowledge
indicating there should be a new function call wrapping the entire buggy string, it fails to identify the correct location to add masks and thus fails to fix this type error. Codex~\cite{codex} is a large language model from OpenAI. Powered by the huge knowledge base stored in the model, Codex identifies that this type error is related to \textit{bytes} types, but it adds checks for the \textit{user} and \textit{password} instead of the entire string, failing to fix this type error. This may be because adding type conversions for variables is much more frequent than that for the entire expression. PyTER~\cite{pyter} is a rule-based approach via dynamic analysis. It fails to find the correct variable inducing the type error, and also cannot introduce the required user-defined type conversion function \textit{to\_bytes}.

\begin{lstlisting}[label=lst:motex, caption=A type error in \textsc{BugsInPy} benchmark, numbers=none]
# Buggy Code: scrapy/scrapy:f042ad
if user:
<@\colorbox{red!30!}{- \quad \quad user\_pass = '\%s:\%s' \% (unquote(user), unquote(password))}@>
<@\colorbox{green!30!}{+ \quad \quad user\_pass = to\_bytes('\%s:\%s' \% (unquote(user), unquote(password)))}@>
      raise ValueError('Port cannot be 0 or less.')
      creds = base64.b64encode(user_pass).strip()
else:
      creds = None
# Patches:
<@\textcolor{lightred}{\# Incorrect Patch provided by CoCoNuT}@>
<@user\_pass = 'Proxy-Authorization'\%(unquote(user), unquote(password))@>
<@\textcolor{lightred}{\# Incorrect Patch provided by AlphaRepair}@>
<@user\_pass = '\%s:\%s' \% (ascii(user), unquote(password))@>
<@\textcolor{lightred}{\# Incorrect Patch provided by Codex}@>
<@if not isinstance(user, bytes):@>
<@\quad \quad user = user.encode('ascii')@>
<@if not isinstance(password, bytes):@>
<@\quad \quad password = password.encode('ascii')@>
<@\textcolor{lightred}{\# Incorrect Patch provided by PyTER}@>
<@if isinstance(creds, bytes):@>
<@\quad \quad creds = str(creds, 'utf-8')@>
<@\textcolor{lightgreen}{\# Correct Patch provided by \tool}@>
<@user\_pass = to\_bytes('\%s:\%s' \% (unquote(user), unquote(password)))@>
\end{lstlisting}

\textbf{\tool.} Before fixing a type error, \tool first mines fix templates from existing type error fixes via hierarchical clustering. 
In the clustering process, \tool can generalize the fix pattern of adding a type conversion for a variable into that of adding a type conversion for an expression. Even though the later fix pattern has low occurrence frequency, \tool can still successfully identify and apply the fix pattern to this type error. Guided by the selected fix template, \tool adds a new function to wrap the original buggy string, and inserts masks for the name of the new function, instead of randomly masking several tokens. As a prompt-based APR approach, \tool also mitigates the problem of introducing user-defined type conversion functions in rule-based approaches like PyTER, since language models can learn from the contexts of the type error. Therefore, \tool can successfully fix the type error.

\section{Approach}\label{sec:approach}
% \label{sec:overview}

\begin{figure*}[t]
    \centering
    \setlength{\belowcaptionskip}{-0.3cm}
    \includegraphics[width = 0.9\textwidth]{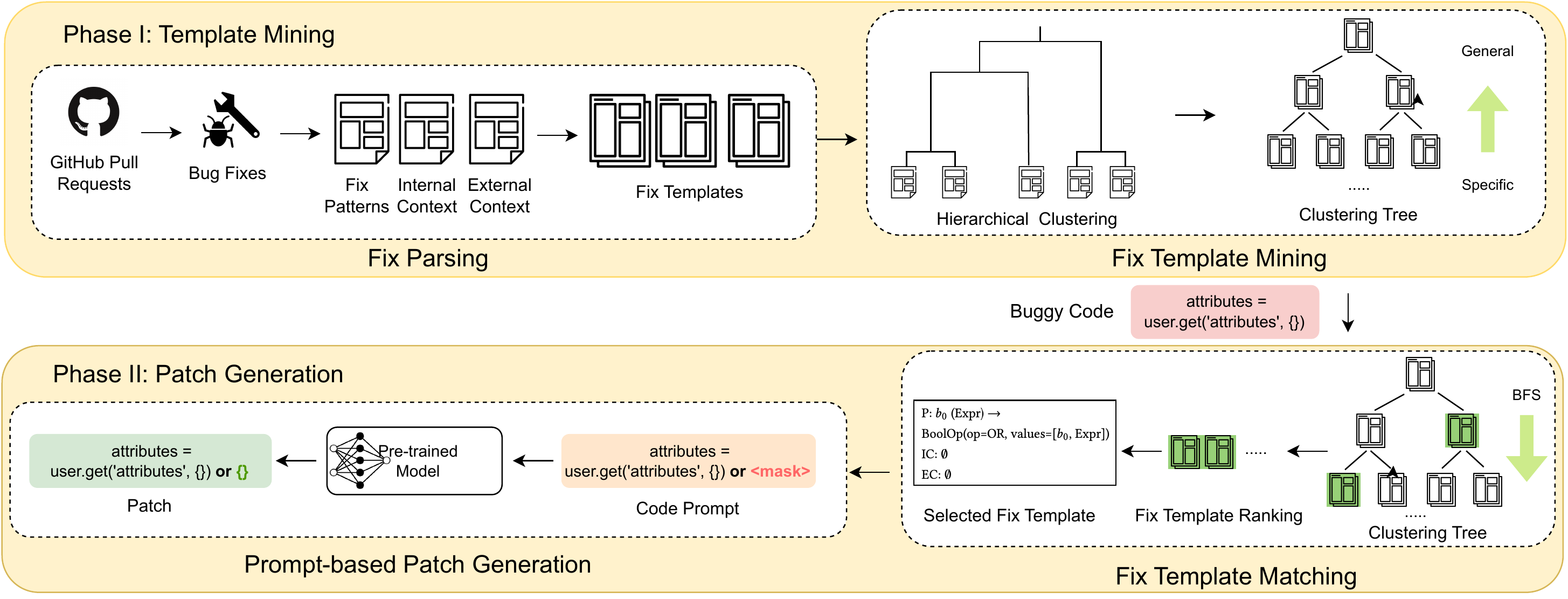}
    \caption{Overview of \tool}
    \label{fig:overview}
\end{figure*}

\tool contains two main phases: the template mining phase and patch generation phase, with the overview shown in Fig.~\ref{fig:overview}.
% We show the overview of \tool in Fig.~\ref{fig:overview}.
In the \textit{template mining} phase, \tool aims at extracting domain-aware fix templates. \tool first parses existing type error fixes into specific fix templates and then employs a novel hierarchical clustering algorithm to abstract and merge them into general fix templates. \tool also organizes the specific to general fix templates into \textit{clustering trees}. In the \textit{patch generation} phase, \tool aims at generating patches for new buggy programs by incorporating prompts with the mined fix templates. Specifically, it selects and ranks the mined fix templates, and applies them on buggy code to automatically generate domain-aware code prompts. The CodeT5~\cite{wang21codet5} model is finally invoked to generate patches by filling the masks in the code prompts.

\subsection{Mining Phase}
The template mining phase mainly contains two stages: fix parsing and fix template mining. The fix parsing stage aims to transform type error fixes into specific fix templates, and the fix template mining stage abstracts and merges parsed specific fix templates into general fix templates via the proposed hierarchical clustering algorithm. We first give formal definitions of fix templates for ease of understanding.

%fix, \tool first parses it into an initial fix template. A fix template contains three components: \textit{fix pattern} indicating the code change, \textit{internal context} indicating the deepest statement that contains the code change, and \textit{external context} indicating other statements in the program. Instead of only including fix patterns in fix templates, we also include internal contexts and external contexts based on the insight that certain bug fix patterns only happen under certain contexts. \tool represents three components of fix templates based on template trees as both code changes and contexts can be represented as trees.

\begin{figure*}[t]
    \centering
    \setlength{\belowcaptionskip}{-0.3cm}
    \includegraphics[width = 0.9\textwidth]{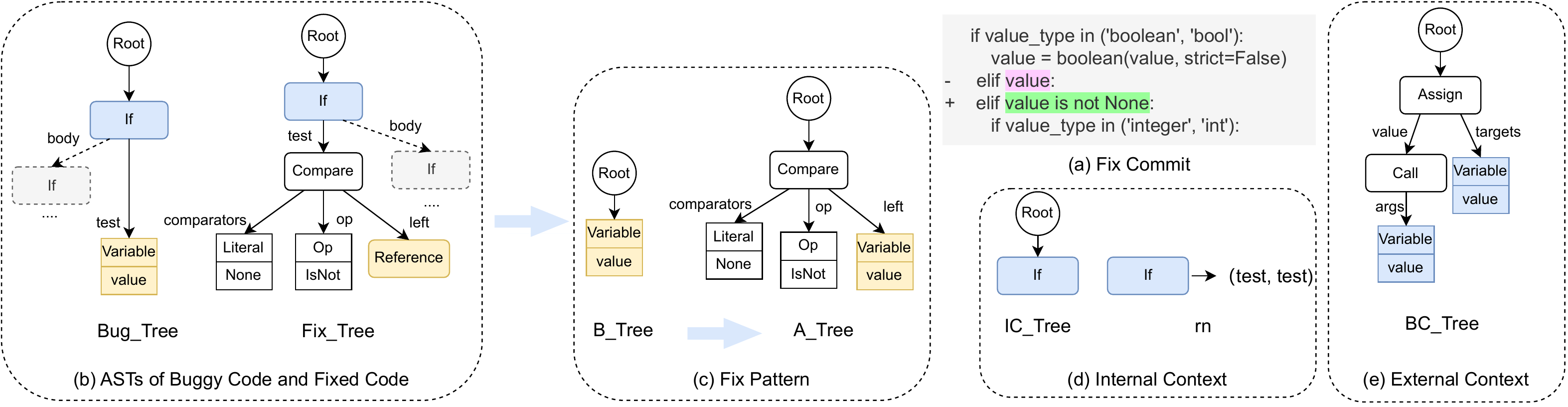}
    \caption{An example of fix parsing process on the fix commit ansible:075c6e.}
    \label{fig:ex}
\end{figure*}

\subsubsection{Definition of Fix Template}
To represent the domain knowledge of where and how to add masks in buggy code for building code prompts, we define fix template as a combination of three parts: \textit{fix pattern}, \textit{internal context} and \textit{external context}. The fix pattern indicates how the buggy code is edited to fix the type error, the internal context pinpoints the locations for applying fix patterns to handle type errors at different levels, and the external context indicates the location of the internal context in the entire buggy program. The three components are all represented based on \textit{template trees} which are defined below.

\definition{3.1.1.1}{Template Tree}
A template tree is a tree ($N$, $E$, $rt$) with nodes $N$, edges $E$ and root node $rt \in N$. An edge is a triple ($n$, $n'$, $r$) where node $n$ is the parent of $n'$ with relation $r$. A node is a quadruple ($bt$, $t$, $v$, $i$) where $bt \in$ \{Variable, Op, Literal, Builtin, Type, Attribute, Expr, Stmt\} is the base type of node, $t$ is the AST node type, $v$ is the value, and $i$ is the id. $bt$, $t$, and $v$ have a special value \textit{ABS} to represent a hole.

We define template trees based on the abstract syntax tree (AST)~\cite{ast} of Python. Keeping the original AST node type $t$, we add a base type $bt$ by re-classifying all original AST node types and attribute types into eight base types, and thus a base type can include multiple AST node types, for example, AST node types \textit{BoolOp}, \textit{BinOp}, and \textit{UnaryOp} belong to the same base type \textit{Expr}. We design base types to obtain a higher level of abstraction than that of ASTs. For instance, the above three AST node types can all be the conditions of \textit{If} statements that serve as guards to prevent type errors. Representing the three AST node types as \textit{Expr} to indicate general conditions help create more general fix templates.

% Based on the definition of template trees, we then
% define fix pattern, internal context and external context.

\definition{3.1.1.2}{Fix Pattern}
A fix pattern is a map $B\_Tree \rightarrow A\_Tree$, where $B\_Tree$ is a template tree of the buggy code, and $A\_Tree$ is a template tree of the fixed code.

\definition{3.1.1.3}{Internal Context}
An internal context is a pair ($IC\_Tree$, $rn$), in which
$IC\_Tree$ is a template tree of the deepest statement where a fix pattern locates, and $rn$ is a map $n \rightarrow (br, ar)$ where $br$ and $ar$ are edge relations, $n \in IC\_Tree.N$ indicates the node where $B\_Tree$ is removed with the edge $(n, B\_Tree.rt, br)$ and $A\_Tree$ is added with the edge $(n, A\_Tree.rt, ar)$.

The internal context is defined to handle edits at different levels. For example, some expression-level edits only modify single expressions in the statements, while other statement-level edits replace the entire statements. Since fix patterns only represent the edits themselves, we use internal contexts to represent the rest parts of the deepest statements for expression-level edits. The internal contexts are empty when the edits are at the statement level.

\definition{3.1.1.4}{External Context}
An external context is a pair ($BC\_Tree$, $AC\_Tree$), where $BC\_Tree$ is a template tree of statements before the internal context and $AC\_Tree$ is a template tree of the statements after the internal context. 

We define external contexts to provide extra location information when $B\_Tree$ in the fix pattern and the internal context are both empty. This usually happens when the fix is about adding
a new statement and does not modify existing buggy code. The \textit{fix template} is a combination of three components including the fix pattern, internal context and external context, defined as below.

\definition{3.1.1.5}{Fix Template}
A fix template is a triple ($P$, $IC$, $EC$), where $P$ ($P \neq \emptyset$) is the fix pattern, $IC$ is the internal context, and $EC$ is the external context. 

%In a fix template, only the fix pattern $P$ represents the edits that we should make to fix a type error, so it cannot be empty. On the contrary, internal contexts and external contexts can be empty since they are used to locate the edit points and match templates. Note that we say a fix pattern or a context is empty if all of their components are empty and we use $\emptyset$ to indicate empty fix patterns and contexts in the entire paper.

We classify fix templates into four categories based on the fix patterns $P$:
\begin{itemize}
    \item \textbf{Add}: $B\_Tree = \emptyset \land A\_Tree \neq \emptyset$
    \item \textbf{Remove}: $B\_Tree \neq \emptyset \land A\_Tree = \emptyset$
    \item \textbf{Insert}: $B\_Tree \neq \emptyset \land A\_Tree \neq \emptyset \land B\_Tree \subset A\_Tree$
    \item \textbf{Replace}: $B\_Tree \neq \emptyset \land A\_Tree \neq \emptyset \land B\_Tree \nsubseteq A\_Tree$
\end{itemize}

Note that there could be more fine-grained classifications under the \textit{Replace} category such as shuffling the order of statements. However, we find that except for the \textit{Insert} category, these cases are really rare (less than 10 cases in the dataset), so we just adopt the general \textit{Replace} category.

\subsubsection{Fix Parsing}
In the fix parsing process, \tool parses all type error fixes into specific fix templates, with an example illustrated in Fig.~\ref{fig:ex}.

\textbf{Parsing Fix Patterns and Internal Contexts.}
Given a fix commit, \tool first extracts the line information of all added and deleted statements, and then walks through the ASTs of buggy code and fixed code to build template trees. To handle edits at different levels, \tool locates the deepest statement-level AST nodes that contain the modified lines, and extracts the corresponding sub-trees in buggy code and fixed code as $Bug\_Tree$ and $Fix\_Tree$, respectively. For example, in the fix commit shown in Fig.~\ref{fig:ex}(a), \tool locates the \textit{If} nodes in the ASTs of buggy code and fixed code, since it is the deepest statement-level AST node containing the edits about variable \textit{value}. Fig.~\ref{fig:ex}(b) illustrates the extracted sub-trees as $Bug\_Tree$ and $Fix\_Tree$. \tool then prunes the same sub-trees shared by $Bug\_Tree$ and $Fix\_Tree$ to leave only the changed part. For example, the bodies of \textit{If} nodes (grey nodes) are pruned and only the conditions remain in Fig.~\ref{fig:ex}(b). 

As the pruned $Bug\_Tree$ and $Fix\_Tree$ contain only the edit, \tool can check whether the edit is at the expression level or the statement level. If $Bug\_Tree$ and $Fix\_Tree$ share the same root node, \tool determines that the edit does not rewrite the entire statement and thus it is at the expression level. Otherwise, \tool can determine that the edit is at the statement level. For instance, $Bug\_Tree$ and $Fix\_Tree$ in Fig.~\ref{fig:ex}(b) have the same root node $If$ (blue nodes), so the edit is at the expression level. For statement-level edits, the internal context is empty. For expression-level edits, \tool creates the internal context by extracting the same nodes shared by $Bug\_Tree$ and $Fix\_Tree$ to form a new template tree $IC\_Tree$, and subtracts $IC\_Tree$ from $Bug\_Tree$ and $Fix\_Tree$ to build $B\_Tree$ and $A\_Tree$ in the fix pattern. The relations of edges that connect $IC\_Tree$ in the internal context and $B\_Tree$ and $A\_Tree$ in the fix pattern are also recorded in the internal context. In the example of Fig.~\ref{fig:ex}, the \textit{If} node is extracted as $IC\_tree$ in the internal context in Fig.~\ref{fig:ex}(d), and the final $B\_Tree$ and $A\_Tree$ constitute fix pattern in Fig.~\ref{fig:ex}(c).

%In the next step, \tool examines whether there are the same nodes shared by two template trees from the paths of root to leaves to determine whether the edits are statement-level or expression-level. If the edits are statement-level, there is no shared nodes and no need to create internal contexts. If the edits are expression-level like Fig.~\ref{fig:ex}, \tool builds a new template tree $IC\_Tree$ for the shared nodes and subtracts the original two template trees $B\_Tree$ and $A\_Tree$ with $IC\_Tree$. In Fig.~\ref{fig:ex}(c), the edit occurs in the condition of \textit{If} nodes and the original two template trees share the \textit{If} node, so \tool extracts the \textit{If} node as well as the node attribute \textit{test} to form the internal context. 

%To determine the category according to the fix pattern, \tool further checks the relationship between $B\_Tree$ and $A\_Tree$. For fix patterns where $B\_Tree$ and $A\_Tree$ are not empty, \tool checks whether $B\_Tree$ is a sub-tree of $A\_Tree$ to detect \textit{Insert} patterns. In Fig.~\ref{fig:ex}(b), \tool finds that $B\_Tree$ becomes a sub-tree of $A\_Tree$ (the old condition becomes an operand of the new condition), so it creates a special reference node in $A\_Tree$ referring to $B\_Tree$. Such references also indicate \textit{Insert} patterns, and no reference indicates \textit{Replace} patterns.

\textbf{Parsing External Contexts.}
\tool identifies statements that locate outside the scope of the internal context but have direct data dependencies with the fix pattern as external contexts. Specifically, it extracts statements that share the same variables with fix patterns before and after the internal context to build $BC\_Tree$ and $AC\_Tree$, respectively. To simplify the fix template abstraction process, \tool also prunes the sub-trees in $BC\_Tree$ and $AC\_Tree$ that do not contain shared variables. For example, in Fig.~\ref{fig:ex}, \tool identifies the statement \textit{value = boolean(value, strict=False)} because it contains the same variable \textit{value} used in the fix pattern. \tool builds a template tree $BC\_Tree$ based on this statement and prunes irrelevant sub-trees such as \textit{strict=False}. Since there is no statement after internal contexts that shares the same variables with fix pattern in Fig.~\ref{fig:ex}(c), $AC\_Tree$ is left empty. The final parsed external context is shown in Fig.~\ref{fig:ex}(e).

\subsubsection{Fix Template Mining}

In the fix template mining process, \tool abstracts and merges the specific fix templates into general fix templates via hierarchical clustering. The rationale of template mining is to ensure the least loss of domain knowledge in fix templates. Based on this rationale, \tool abstracts or merges the two most similar fix templates each time, and organizes specific to general fix templates as clustering trees. To measure the similarity between two fix templates, we define two kinds of similarity metrics: \textit{value distance} and \textit{structural distance}. The \textit{structural distance} measures the ratio of nodes in two template trees that have the same type regardless of values (type matching), while the \textit{value distance} measures the ratio of nodes in two template trees that have the same types and values (value matching).

\definition{3.1.3.1}{Fix Pattern Distances}
The value distance $d_p$ and structural distance $sd_p$ between two template trees in fix patterns are defined as
\begin{equation}
    d_p(t_1, t_2) = 1 - \frac{VM_p(t_1.rt, t_2.rt)}{\text{Num}(t_1) + \text{Num}(t_2)}
    \nonumber
\end{equation}
\begin{equation}
    sd_p(t_1,t_2) = 1 - \frac{TM_p(t_1.rt, t_2.rt)}{\text{Num}(t_1) + \text{Num}(t_2)},
    \nonumber
\end{equation}
% , 
where $Num(t)$ indicates the number of nodes in the template tree $t$, and ValueMatch $VM_p$ and TypeMatch $TM_p$ are defined as
\begin{equation}
    VM_p(n_1, n_2) = \left\{ 
    \begin{array}{ll}
        0 & n_1 \neq n_2 \\ 
        2 + \sum_{i \in \text{child}(n_1, n_2)} VM_p(n_1^i, n_2^i) & \text{otherwise}
    \end{array}
    \right.
    \nonumber
\end{equation}
\begin{equation}
    TM_p(n_1, n_2) = \left\{ 
    \begin{array}{ll}
        0 & n_1.t \neq n_2.t \\ 
        2 + \sum_{i \in \text{child}(n_1, n_2)} TM_p(n_1^i, n_2^i) & \text{otherwise}
    \end{array}
    \right.
    \nonumber
\end{equation}

\definition{3.1.3.2}{Context Distances}
The value distance $d_c$ and structural distance $sd_c$ between two template trees in contexts are defined as
\begin{equation}
    d_c(t_1, t_2) = 1 - \frac{\text{MAX}(\text{LeafNode}(t_1), \text{LeafNode}(t_2), VM_c)}{\text{Num}(t_1) + \text{Num}(t_2)}
    \nonumber
\end{equation}
\begin{equation}
    sd_c(t_1, t_2) = 1 - \frac{\text{MAX}(\text{LeafNode}(t_1), \text{LeafNode}(t_2), TM_c)}{\text{Num}(t_1) + \text{Num}(t_2)},
    \nonumber
\end{equation}
% , 
where MAX($a$, $b$, $c$) pairs the elements in $a$ and $b$, and finds the highest similarity $c$ achieved by the pairs, and returns the number of pairs, $Num(t)$ indicates the number of nodes in the template tree $t$, and $LeafNode(t)$ returns the leaf node set of a template tree $t$. The ValueMatch $VM_c$ and TypeMatch $TM_c$ are defined as
\begin{equation}
    VM_c(n_1, n_2) = \left\{ 
    \begin{array}{ll}
        0 & n_1 \neq n_2 \\ 
        2 + VM_c(n_1.parent, n_2.parent) & \text{otherwise}
    \end{array}
    \right.
    \nonumber
\end{equation}
\begin{equation}
    TM_c(n_1, n_2) = \left\{ 
    \begin{array}{ll}
        0 & n_1.t \neq n_2.t \\ 
        2 + TM_c(n_1.parent, n_2.parent) & \text{otherwise}
    \end{array}
    \right.
    \nonumber
\end{equation}

To calculate the distances of fix patterns, we adopt a top-down methodology. We start with the root node and require two nodes to be type-matching or value-matching before we compare their child nodes. To calculate the distances of contexts, we adopt a bottom-up methodology. We start with the leaf nodes and require two nodes to be type-matching or value-matching before we compare their parent nodes. Such a difference is caused by the functionality of fix patterns and contexts. The template trees in the fix patterns are used to generate patch code, so based on the definition of ASTs the children nodes are meaningful only if their parent nodes are determined. On the contrary, the template trees in the contexts are used to match the locations that fix patterns should apply instead of generating code, so the children nodes contain more specific location information than the parent nodes. For example, in Fig.~\ref{fig:ex}(d), even if we remove the node \textit{Assign}, it can still match the original statement through \textit{Call}, but if we remove the node \textit{Variable}, it can match more general statements that have no direct data dependency with fix pattern in Fig.~\ref{fig:ex}(b).

\textbf{Template Abstraction.} \tool does not abstract the whole fix template, instead, it abstracts one component, i.e., fix pattern, internal context or external context, each time. \tool abstracts the two similar components through a process named \textit{Abstract}. Fig.~\ref{fig:abs_pattern} and Fig.~\ref{fig:abs_context} formally present the processes of \textit{Abstract} on fix patterns and contexts, respectively.

The abstraction of fix patterns and contexts follows the aforementioned top-down and bottom-up methodology, respectively. Generally, there could be four cases when abstracting the template node $a$ and $b$ from two similar template trees:
\begin{itemize}
    \item \textbf{Same Node}: $a$ and $b$ are exactly the same, and they can be reserved for the generalized fix template.
    \item \textbf{Value Abstraction}: $a$ and $b$ have the same types but different values. \tool creates a node with the same type and set the value as a special \textit{ABS} token to indicate a hole.
    \item \textbf{Type Abstraction}: $a$ and $b$ have the same base types but different types and values. \tool creates a node with the same base type, and sets the type and value as a special \textit{ABS} token to indicate a hole.
    \item \textbf{Node Removal}: $a$ and $b$ have no common attributes. \tool directly removes the two nodes.
\end{itemize}
\tool also prunes all child nodes in Type Abstraction and Node Removal, because the change of types for an AST node disables the functionality of its original child nodes.

%For fix templates under \textit{Insert} category, \tool masks the sub-tree in $A\_Tree$ that is the same with $B\_Tree$ with a single reference node before abstraction. This is to avoid different abstraction results between $B\_Tree$ and the sub-tree in $A\_Tree$.
\begin{figure}[t]
    \centering
    \begin{equation}
    \begin{split}
    \textbf{Abstract}(n_1(C_{r_1}^1,...,C_{r_m}^1), n_2(C_{r_1}^2,...,C_{r_n}^2)) = & \\
    \left\{
    \begin{array}{ll}
        n_1(O_{r_1},...,O_{r_k}) & \textbf{if} \ n_1 = n_2 \\
         & \textbf{where} \ k = \text{min}(m,n), \\
         & p = \text{min}(\text{len}(C^1_{r_i}), \text{len}(C^2_{r_i})), \ O_{r_i} = \{O^1_{r_i},...,O^p_{r_i}\},\\
         & O^j_{r_i} = \textbf{Abstract}(C^{1j}_{r_i}, C^{2j}_{r_i}) \ \forall j \in [1,p] \\
         & \text{(Same Node)} \\
        o(O_{r_1},...,O_{r_k}) & \textbf{if} \ n_1.v \neq n_2.v \land n_1.t = n_2.t \land n_1.bt = n_2.bt \\
        & \textbf{where} \ o.v = ABS,\ o.t = n_1.t,\ o.bt = n_1.bt,\\
        & k = \text{min}(m,n),\ p = \text{min}(\text{len}(C^1_{r_i}), \text{len}(C^2_{r_i})), \\
        & O_{r_i} = \{O^1_{r_i},...,O^p_{r_i}\},\\
         & O^j_{r_i} = \textbf{Abstract}(C^{1j}_{r_i}, C^{2j}_{r_i}) \ \forall j \in [1,p] \\
         & \text{(Value Abstraction)} \\
         o & \textbf{if} \ n_1.t \neq n_2.t \land n_1.bt = n_2.bt \\
         & \textbf{where} \ o.v = ABS,\ o.t = n_1.bt,\ o.bt = n_1.bt \\
         & \text{(Type Abstraction)} \\
         \emptyset & \textbf{otherwise} \\
         & \text{(Node Removal)}
    \end{array}
    \right. & \\
    \textbf{where} \ n_1, n_2 \in A\_Tree.N \ \textbf{or} \ n_1, n_2 \in B\_Tree.N, C^j_{r_i} = \{C^{jt}_{r_i}\} & \\
     \textbf{s.t.} \ \text{Edge}(n^j, C^{jt}_{r_i}, r_i) \in A\_Tree.E \
    \textbf{or} \ \text{Edge}(n^j, C^{jt}_{r_i}, r_i) \in B\_Tree.E &
    \end{split}
    \nonumber
\end{equation}
\begin{equation}
    \begin{split}
        \textbf{Abstract}(P(a_1, b_1), P(a_2, b_2)) = 
        P(a, b)& \\
        \textbf{where} \ a.rt = \textbf{Abstract}(a_1.rt, a_2.rt) & \\
        b.rt = \textbf{Abstract}(b_1.rt, b_2.rt) & 
    \end{split}
    \nonumber
\end{equation}
    \caption{The process of Abstraction for fix patterns. }
    \label{fig:abs_pattern}
\end{figure}

\begin{figure}[t]
    \centering
    \begin{equation}
    \begin{split}
    \textbf{Abstract}(n_1(P^1_{r_1}), n_2(P^2_{r_2})) = & \\
    \left\{
    \begin{array}{ll}
        n_1(O_r) & \textbf{if} \ n_1 = n_2\\
         & \textbf{where} \ O_r = \textbf{Abstract}(P^1_{r_1}, P^2_{r_2}) \ \text{if} \ r_1 = r_2,\\
         & O_r = \emptyset \ \text{if} \ r_1 \neq r_2 \\
         & \text{(Same Node)} \\
         %n_1 & \textbf{if} \ n_1 = n_2 \land r_1 \neq r_2 \\
         %& \text{(Same Node)} \\
        o(O_r) & \textbf{if} \ n_1.v \neq n_2.v \land n_1.t = n_2.t \land n_1.bt = n_2.bt \\
        & \textbf{where} \ o.v = ABS, \ o.t = n_1.t,\ o.bt = n_1.bt,\\
        & O_r = \textbf{Abstract}(P^1_{r_1}, P^2_{r_2}) \ \text{if} \ r_1 = r_2, \\
        & O_r = \emptyset \ \text{if} \ r_1 \neq r_2 \\
         & \text{(Value Abstraction)} \\
         %o & \textbf{if} \ n_1.v \neq n_2.v \land n_1.t = n_2.t \land n_1.bt = n_2.bt \land r_1 \neq r_2 \\
         %& \textbf{where} \ o.v = ABS \land o.t = n_1.t \land o.bt = n_1.bt\\
         %& \text{(Value Abstraction)} \\
         o & \textbf{if} \ n_1.t \neq n_2.t \land n_1.bt = n_2.bt \\
         & \textbf{where} \ o.v = ABS,\ o.t = n_1.bt,\ o.bt = n_1.bt \\
         & \text{(Type Abstraction)} \\
         \emptyset & \textbf{otherwise} \\
         & \text{(Node Removal)}
    \end{array}
    \right. & \\
    \textbf{where} \ n_1, n_2 \in IC\_Tree.N \ \textbf{or} \ n_1, n_2 \in BC\_Tree.N \\ \textbf{or} \ n_1, n_2 \in AC\_Tree.N,
    \text{Edge}(n^j, P^{j}_{r_j}, r_j) \in IC\_Tree.E \\
    \textbf{or} \ \text{Edge}(n^j, P^{j}_{r_j}, r_j) \in BC\_Tree.E \
    \textbf{or} \ \text{Edge}(n^j, P^{j}_{r_j}, r_j) \in AC\_Tree.E & 
    \end{split}
    \nonumber
\end{equation}
\begin{equation}
    \begin{split}
        \textbf{Abstract}(IC(a_1), IC(a_2), Pairs(c)) = 
        IC(a)& \\
        \textbf{where} \ \text{LeafNode}(a) = \{\textbf{Abstract}(p^i_1, p^i_2)\} \ \forall p^i \in c& \\
    \end{split}
    \nonumber
\end{equation}
\begin{equation}
    \begin{split}
        \textbf{Abstract}(EC(a_1, b_1), EC(a_2, b_2), Pairs(c_1, c_2)) = 
        EC(a, b)& \\
        \textbf{where} \ \text{LeafNode}(a) = \{\textbf{Abstract}(p^i_1, p^i_2)\} \ \forall p^i \in c_1& \\
        \text{LeafNode}(b) = \{\textbf{Abstract}(p^i_1, p^i_2)\} \ \forall p^i \in c_2 &
    \end{split}
    \nonumber
\end{equation}
    \caption{The process of Abstraction for both internal context and external context.}
    \label{fig:abs_context}
\end{figure}

\textbf{Mining Fix Templates via Hierarchical Clustering.}
\begin{algorithm}[t]
\caption{Fix Template Mining}
\label{alg:mining}
\begin{algorithmic}[1]
\Require
A set of parsed specific fix templates, $T$
\Ensure
Mined fix templates, $CT$
\State $CT \leftarrow T$
\While{isChanged($CT$)}
\State $D_{p}, D_{IC}, D_{EC} \leftarrow$ CalculateValueDistances($CT$)
\State $SD_{p}, SD_{IC}, SD_{EC} \leftarrow$ CalculateStructuralDistances($CT$)
\State $CT \leftarrow$ Deduplicate($CT$)
%\State // Abstract and merge external contexts
\State $clusters \leftarrow$ selectClusters($CT$, $D_P$, $D_{IC}$, $D_{EC}$, $SD_{EC}$) 
\For{$c \in clusters$}\Comment{Handle external contexts}
\State $t_1, t_2 \leftarrow$ argmin($c$, $D_{EC}$); $nt \leftarrow t_1$
\State $pairs \leftarrow$ getPairs($D_{EC}$($t_1, t_2$))
\State $nt.EC \leftarrow$ Abstract($t_1.EC, t_2.EC, pairs$, Context)
\State $CT \leftarrow CT - \{t_1, t_2\} + \{nt\}$
\State $t_1.parent, t_2.parent \leftarrow nt$
\EndFor
\State continue if isChanged($CT$)
%\State // Abstract internal contexts
\State $clusters \leftarrow$ selectClusters($CT$, $D_P$, $D_{IC}$, $SD_{IC}$)
\For{$c \in clusters$}\Comment{Handle internal contexts}
\State $t_1, t_2 \leftarrow$ argmin($c$, $D_{EC}$); $nt_1 \leftarrow t_1$; $nt_2 \leftarrow t_2$
\State $pairs \leftarrow$ getPairs($D_{IC}$($t_1, t_2$))
\State $nt_1.IC, nt_2.IC \leftarrow$ Abstract($t_1.IC, t_2.IC, pairs$, Context)
\State $CT \leftarrow CT - \{t_1, t_2\} + \{nt_1, nt_2\}$
\State $t_1.parent \leftarrow nt_1$; $t_2.parent \leftarrow nt_2$
\EndFor
\State continue if isChanged($CT$)
%\State // Abstract fix patterns
\State $clusters \leftarrow$ selectClusters($CT$, $D_P$, $SD_{P}$)
\For{$c \in clusters$}\Comment{Handle fix patterns}
\State $t_1, t_2 \leftarrow$ argmin($c$, $D_{P}$); $nt_1 \leftarrow t_1$; $nt_2 \leftarrow t_2$
\State $nt_1.P, nt_2.P \leftarrow$ Abstract($t_1.P, t_2.P$, Pattern)
\State $CT \leftarrow CT - \{t_1, t_2\} + \{nt_1, nt_2\}$
\State $t_1.parent \leftarrow nt_1$; $t_2.parent \leftarrow nt_2$
\EndFor
\EndWhile

\end{algorithmic}

\end{algorithm}

With the above-mentioned similarity metrics and abstraction processes, \tool selects similar fix templates and merges them via hierarchical clustering to build clustering trees. We give the definition of the clustering tree as follows.

\definition{3.1.3.3}{Clustering Tree}
A clustering tree is a tree $(T, E, rt)$ with fix template set $T$, edges $E$ and root fix template $rt \in T$. An edge is a pair $(t, t')$ where fix template $t$ is the parent of fix template $t'$, indicating that $t$ is directly abstracted from $t'$. 

To ensure the least loss of domain knowledge, \tool follows two strategies in the mining process. First, \tool follows the priority order of ``external context $>$ internal context $>$ fix pattern'' when selecting component pairs for abstraction, ensuring that abstraction of fix patterns happens only if no external context pairs and internal context pairs can be abstracted. Second, \tool prefers value abstraction to type abstraction, so it prioritizes component pairs, i.e., fix pattern pairs, internal context pairs or external context pairs, with a structural distance at 0.

We present the hierarchical clustering algorithm of \tool in Alg.~\ref{alg:mining}. At the beginning of each iteration, \tool calculates the distances of three components for every two fix templates (lines 3$\sim$4). \tool then removes duplicated fix templates. Based on the calculated distances, \tool first handles external context pairs (lines 7 $\sim$ 13), then internal context pairs (lines 16 $\sim$ 22), and finally fix patterns (lines 25 $\sim$ 30). If any abstraction or merge happens, the current iteration will be terminated and a new iteration will begin (lines 14 and 23).  

When handling external context pairs, \tool groups the fix templates with the same internal contexts and fix patterns into different clusters (line 6). When handling internal context pairs, \tool groups the fix templates with the same fix patterns into different clusters (line 15). When handling fix patterns, all fix templates are grouped into one single cluster (line 24). This ensures that only fix templates in the same cluster can be abstracted and merged into more general fix templates under the priority order. For each cluster, \tool selects the certain components with the lowest distance in two fix templates (lines 8, 17, 26), and abstracts them into more general components in each iteration (lines 10, 19, 27). The selection has two stages. In the first stage, only components with a structural distance of 0 in the fix templates are considered to prioritize value abstraction. In the second stage, when no such component exists, the rest components are considered. Note that there could be trivial abstractions such as removing all nodes for a template tree so that an empty template tree can represent any code. As empty fix patterns provide no domain knowledge for patch generation, \tool does not select two fix patterns whose distance and structural distance are both at 1 in the fix templates for further abstraction.

At the end of each iteration, the new fix templates are included in the set, and the old fix templates are removed from the set (lines 11, 20, 28). The relationships between the new fix templates and the old fix templates are recorded in the clustering tree (lines 12, 21, 29). The merge of fix templates happens only when handling external contexts since the fix patterns and internal contexts of fix templates are already required to be the same at this time. \tool also records the number of instances each fix template represent in the training set to facilitate fix template ranking in the next phase. When the template mining process completes, clustering trees that contain specific to general fix templates are generated, facilitating the next patch generation phase.

%In the mining process, \tool creates generalized fix templates by abstracting each component of old fix templates. The merge of fix templates only happens in the external context abstraction since it already requires internal contexts and fix patterns to be the same. In each abstraction or merge, \tool maintains a clustering tree and sets the new generalized fix templates as the parents of old fix templates (lines 13, 23, 32 in Alg.~\ref{alg:mining}). 

\subsection{Patch Generation Phase}
The patch generation phase of \tool mainly contains two processes: fix template matching and prompt-based patch generation. The fix template matching process aims to select and rank appropriate fix templates that could be applied to the buggy program. The prompt-based patch generation process aims to generate candidate patches by applying selected fix templates to generate code prompts and invoking code pre-trained models for mask prediction.

\subsubsection{Fix Template Matching}\label{sec:match}
In the fix template matching process, \tool selects matched fix templates on clustering trees via Breadth-First Search (BFS) and then ranks fix templates with frequency and abstraction ratio.

\textbf{Selecting Fix Templates.} Given a buggy program, \tool parses the bug lines into a template Tree $Bug\_Tree$, and the contexts before and after the bug lines into template trees $BBug\_Tree$ and $ABug\_Tree$, respectively. \tool compares the triple ($Bug\_Tree$, $BBug\_Tree$, $ABug\_Tree$) with fix templates in the clustering trees to find the appropriate fix templates. We define the following rules to check whether a buggy program matches a fix template.

\definition{3.2.1.1}{Template Node Match} For two template nodes $a$ and $b$, $a$ matches $b$ if $a.value$ matches $b.value$ and $(a.t, a.bt)$ matches $(b.t, b.bt)$. $a.value$ matches $b.value$ if $b.value = ABS \lor a.value = b.value$. $(a.t, a.bt)$ matches $(b.t, b.bt)$ if $a.bt = b.bt \land (a.bt = b.t \lor a.t = b.t)$.

\definition{3.2.1.2}{Template Tree Match} For two template trees $A$ and $B$, $A$ matches $B$ if there is a node $a \in A.N$ where $a$ matches to $B.rt$ and there exists node maps $\{ac_1 \rightarrow bc_1, ..., ac_n \rightarrow bc_n\}$ where $ac_i$ matches $bc_i$, $\{ac_1, ..., ac_n\} \subseteq a.children$, $ac_{i+1}.id > ac_i.id$, and $\{bc_1,...,bc_n\} =$ $B.rt.children$. $A$ always matches $B$ if $B = \emptyset$.

\definition{3.2.1.3}{Fix Template Match} For a buggy program ($Bug\_Tree$, $BBug\_Tree$, $ABug\_Tree$) and a fix template ($P$, $IC$, $EC$), the buggy program matches the fix template if $BBug\_Tree$ matches $EC.BC\_Tree$, $ABug\_Tree$ matches $EC.A\_Tree$ and $Bug\_Tree$ matches Concat($IC.IC\_Tree$, $P.B\_Tree$,$IC.rn$), where Concat($a$, $b$, $rn$) indicates concatenating template tree $b$ to template tree $a$ with edge ($n$, $b.rt$, $rn[n].br$).

With the above rules, \tool starts with the root fix template (most general fix templates) of each clustering tree and walks through the clustering tree via bread-first search (BFS) until it finds the deepest fix template (most specific fix templates) matched by the buggy program. These fix templates are collected to be ranked in the next step.

\textbf{Ranking Fix Templates.} \tool ranks the fix templates before applying them to the buggy program. To provide the most domain knowledge for pre-trained models in the patch generation process, \tool utilizes a two-step strategy to prioritize fix templates.

\tool groups the fix templates with the same concatenated template tree of $IC\_Tree$ and $B\_Tree$. These fix templates provide different fix solutions for the same buggy pattern. \tool ranks fix templates in one group based on the number of training instances they represent because a larger number indicates that the fix template is used more frequently on the given buggy program. \tool then ranks the groups based on the abstraction ratio of $A\_Tree$ of the first fix template in each group. The abstraction ratio of a template tree is defined by the ratio of nodes whose values or types are $ABS$ tokens. A higher abstraction ratio of $A\_Tree$ indicates less domain knowledge associated, so that code pre-trained models need to predict more information before they can generate complete candidate patches. For example, an abstraction ratio of 1.0 indicates the fix template actually is a huge hole and there is no domain knowledge assisting the generation of patches. Therefore, \tool prioritizes the groups with a lower abstraction ratio to include more domain knowledge in the patch generation process.

\subsubsection{Prompt-based Patch Generation}\label{sec:metric}
In the process, \tool applies ranked fix templates on the buggy program and generates code prompts. CodeT5 model is then invoked to fill the masks in code prompts and generate candidate patches.

\textbf{Applying Fix Templates.} For each selected fix template, \tool completes the $A\_Tree$ in its fix pattern by adding dummy AST nodes, i.e., AST nodes with values of $ABS$ tokens, as placeholders, because some child AST nodes are removed in the fix template mining process. \tool then replaces the sub-tree AST of the buggy program that matches $B\_Tree$ with the completed $A\_Tree$, and converts the modified AST of the buggy program to code prompts. Code prompts are source code that contains \textit{ABS} tokens as masks to be predicted by code pre-trained models.

\textbf{Generating Patches.} Most code pre-trained models are trained to predict masks in source code, thus they can naturally be used to predict the value of $ABS$ tokens in the code prompts. In this paper, we choose CodeT5~\cite{wang21codet5} as the code pre-trained model in the patch generation process, since it is specially designed for the code generation task~\cite{wang21codet5}. When generating patches, \tool replaces the \textit{ABS} tokens in the code prompt with ordered mask tokens used in CodeT5, e.g., <extra\_id\_0>, ..., <extra\_id\_99>. \tool then invokes CodeT5 to predict tokens for each mask. The predicted values for the masks are filled into the code prompts to generate candidate patches.

\textbf{Validating Patches.} \tool adopts the classic generate-and-validate methodology in patch generation. For the generated patches, \tool first filters out those with syntax errors, and then runs the test suite on each patch to find plausible patches, i.e., those can successfully pass all test cases. Plausible patches are further examined by the authors to identify correct patches, i.e., those are semantically identical to the developer patch when ignoring I/O side effects such as messages in \textit{print} statements.

\section{Experimental Design}\label{sec:setup}

\subsection{Dataset}\label{sec:dataset}

\textbf{Training Set.} Following previous work~\cite{pyter}, we build a dataset for the fix template mining process of \tool and the training of baselines. We collect 8,722 merged pull requests from GitHub that contain the term ``fix type error''. We extract the fixes from the commits in collected pull requests. We remove the overlong commits that contain more than 50 lines of modified code. Finally, we get 10,981 fixes to form the training set.

\textbf{Benchmarks.} Following previous work~\cite{pyter}, we use two benchmarks \bugsinpy~\cite{bugsinpy} and \typebugs~\cite{pyter}. The two benchmarks initially separate type errors by commits, but we find that a single commit can also involve more than one type errors in different locations. To avoid the correct fix of one type error being hidden by another type error, we further split the commits that contain two or more type errors into multiple ones. We also remove the duplicated type errors, i.e., those that have the same commit signatures, in two benchmarks. Finally, we get 54 type errors from \bugsinpy and 109 type errors from \typebugs.

\subsection{Baselines}
We compare \tool with the following four baselines.

\textbf{PyTER.} PyTER~\cite{pyter} is a rule-based APR approach designed for repairing Python type errors. It has nine pre-defined templates and several rules to synthesize templates to generate candidate patches. 

\textbf{AlphaRepair.} AlphaRepair~\cite{xia22less} is the state-of-the-art prompt-based approach for general-purpose APR. It masks tokens in the buggy code based on some general prompt templates and invokes code pre-trained models to generate patches.

\textbf{CoCoNuT.} CoCoNuT~\cite{Lutellier20coconut} is an NMT-based approach for general-purpose APR. It translates the buggy code into candidate patches. 

\textbf{Codex.} Codex~\cite{codex} is a large GPT model fine-tuned on publicly available code from GitHub. It is designed by OpenAI and used to power GitHub Copilot~\cite{copilot} service.

\subsection{Metrics}
We adopt the commonly used \textbf{Correct} and \textbf{Plausible} metrics in previous work~\cite{Lutellier20coconut,jiang21cure,pyter,xia22less} to evaluate the performance of \tool on repairing type errors. Besides, we add a new metric named \textbf{Template Coverage} to evaluate the number of developer patches covered by the fix templates mined by \tool and pre-defined ones from PyTER. \textbf{Template Coverage} is defined by the ratio of bugs whose developer patch matches a fix template of an approach.

\begin{table}[t]
    \centering
    \caption{Evaluation results of \tool compared with three baselines. Results are presented in the Correct/Plausible format. Fix rate is the ratio of correct patches.}
    \scalebox{0.8}{
    \begin{tabular}{cc|ccccc}
    \toprule
        \multicolumn{7}{c}{\g \textbf{\textsc{TypeBugs}}} \\
        \midrule
        \textbf{Project} & \textbf{\#B}& \textbf{\tool} & \textbf{PyTER} & \textbf{Codex} & \textbf{AlphaRepair} & \textbf{CoCoNuT}\\
        \midrule
        airflow & 14 & 9/9 & 4/4 & 7/7 & 1/6 & 0/4 \\
        beets & 1 & 0/0 & 0/1 & 0/0 & 0/0  & 0/0 \\
        core & 9 & 7/7 & 5/7 & 4/5 & 4/4  & 2/3 \\
        kivy & 1 & 0/0 & 0/1 & 0/0 & 0/1  & 0/1 \\
        luigi & 2 & 0/2 & 0/0 & 0/2 & 1/2 & 0/0 \\
        numpy & 3 & 0/3 & 0/2 & 0/1 & 0/2  & 0/0 \\
        pandas & 48 & 21/32 & 17/27 & 18/19 & 11/22  & 3/10\\
        rasa & 2 & 2/2 & 0/0 & 2/2 & 0/0  & 0/0 \\
        requests & 4 & 4/4 & 4/4 & 2/2 & 0/1 & 0/0 \\
        rich & 4 & 2/3 & 0/1 & 1/1  &  0/0 & 0/0 \\
        salt & 8 & 5/8 & 5/5 & 4/5 & 1/5 & 0/2 \\
        sanic & 2 & 0/0 & 2/2 & 0/0 & 0/0 & 0/0 \\
        scikit-learn & 7 & 2/3 & 2/3  & 1/2 & 0/0 & 0/0 \\
        tornado & 1 & 0/0 & 1/1 & 0/0 & 0/0  & 0/0 \\
        Zappa  & 3 & 3/3 & 1/1 & 0/0 & 1/3 & 0/1 \\
        \midrule
        \textbf{Total} & 109 & \textbf{55/76} & 41/59  & 39/46 &   19/46 & 5/21\\
        \textbf{Fix Rate (\%)}& - & \textbf{50.5} & 37.6  & 35.8 &   17.4 & 4.6\\
        \midrule
        \multicolumn{7}{c}{\g \textbf{\textsc{BugsInPy}}} \\
        \midrule
        \textbf{Project} & \textbf{\#B} & \textbf{\tool} & \textbf{PyTER} & \textbf{Codex} & \textbf{AlphaRepair} & \textbf{CoCoNuT} \\
        \midrule
        ansible & 1 & 0/0 & 0/0 & 0/0 & 0/0 & 0/0 \\
        fastapi & 1 & 1/1 & 0/0 & 1/1 & 0/0 & 0/0 \\
        keras & 7 & 4/6 & 1/1 & 0/3 & 0/4 & 0/4 \\
        luigi & 7 & 4/5 & 3/5 & 3/3 & 0/0  & 0/0 \\
        pandas & 19 & 4/13 & 4/6 & 2/6 & 3/10 & 3/8 \\
        scrapy & 12 & 10/11 & 5/7 & 10/12 & 1/4 & 2/4 \\
        spacy & 1 & 0/1 & 0/1 & 0/0 & 0/1 & 0/1\\
        tornado & 2 & 1/1 & 1/1 & 1/1 & 0/1 & 0/1 \\
        youtube-dl & 4 & 2/3 & 1/1 & 0/2 & 1/1  & 1/1 \\
        \midrule
        \textbf{Total} & 54 & \textbf{26/41} & 15/22 & 17/28 & 5/21 & 6/19\\
        \textbf{Fix Rate (\%)} & - & \textbf{48.1} & 27.8 & 31.5 &  9.3 & 11.1\\
    \bottomrule
    \end{tabular}}
    \label{tab:main_res}
\end{table}

\subsection{Implementation}
The entire framework of \tool is implemented using Python, which contains more than 10,000 lines of code. We adopt the CodeT5-base~\cite{codet5base} model to predict the masks in code prompts and generate candidate patches. For PyTER, AlphaRepair and CoCoNuT, we directly use the replication packages released by the authors and re-implement them on our task. We train CoCoNuT with its original training set and the training set we collected to adapt it to fix Python type errors. Since Codex is not publicly available, we use the public API~\cite{codexapi} of engine \textit{code-davinci-002} provided by OpenAI to query it with prompts. We use a similar prompt from previous work~\cite{xia22practical}. The only difference is that we use three examples instead of two at the beginning of the prompt and only mask the buggy line to maximize the performance of Codex. We make all other settings consistent with previous work~\cite{pyter, xia22less,jiang21cure,Lutellier20coconut,xia22practical}. All experiments are conducted on a Linux machine (Ubuntu 20.04) with two Intel Xeon@2.20GHZ CPUs, one NVIDIA A100-SXM4-40GB GPU and 256GB RAM.
\section{Evaluation}\label{sec:eval}
In this section, we evaluate the performance of \tool on the following three research questions:
\begin{itemize}
    \item \textbf{RQ1}: How effective is \tool to fix type errors?
    \item \textbf{RQ2}: How capable is \tool to mine fix templates from existing bug fixes?
    \item \textbf{RQ3}: When does \tool fail to fix type errors?
\end{itemize}

\subsection{RQ1: Effectiveness of \tool}

To evaluate the effectiveness of \tool on repairing type errors, we compare \tool with state-of-the-art rule-based APR approaches and learning-based APR approaches. Table~\ref{tab:main_res} presents the performance of \tool along with baseline approaches on two benchmarks \textsc{TypeBugs} and \textsc{BugsInPy}.

\textbf{Comparison with Rule-based Approach.} As can be seen in Table~\ref{tab:main_res}, \tool can successfully fix 55 type errors in \textsc{TypeBugs} and 26 type errors in \textsc{BugsInPy}, outperforming rule-based approach PyTER by 14 and 11 type errors, respectively.  We attribute the improvement of \tool to the higher coverage of fix templates mined from existing type error fixes and the generated domain-aware code prompts. Furthermore, we analyze the unique type errors that \tool and PyTER can fix in two benchmarks and present the results in Table~\ref{tab:coverage}. We find that \tool obtains 24 and 16 unique type error fixes in \textsc{TypeBugs} and \textsc{BugsInPy}, respectively, while PyTER only obtains 10 and 5 unique type error fixes in \textsc{TypeBugs} and \textsc{BugsInPy}, respectively. This further demonstrates the effectiveness of \tool when compared with PyTER.

\begin{figure}[t]
    \centering
    \subfigure[\textsc{TypeBugs}]{
    \begin{minipage}[t]{0.47\linewidth}
    \centering
    \includegraphics[width = 1.0\textwidth]{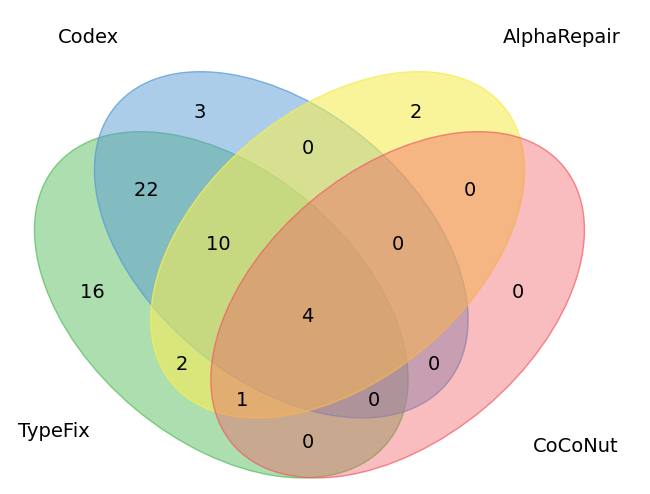}
    \label{fig:typebugs_venn}
    \end{minipage}
    }
    \subfigure[\textsc{BugsInPy}]{
    \begin{minipage}[t]{0.47\linewidth}
    \includegraphics[width = 1.0\textwidth]{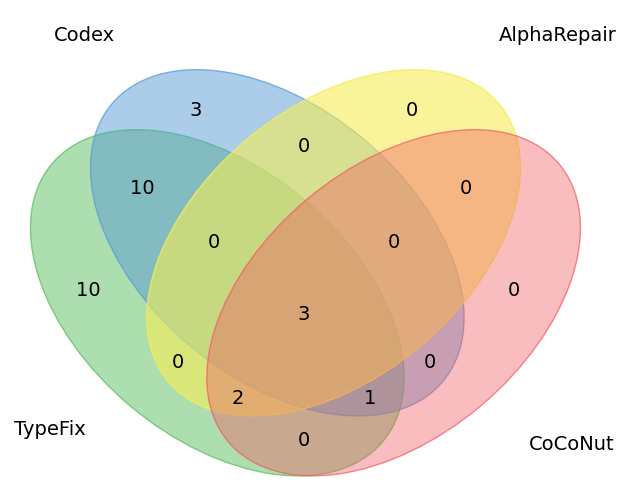}
    \label{fig:bugsinpy_venn}
    \end{minipage}
    }
    \caption{Venn diagram of correct patches provided by learning-based APR approaches.}
    \label{fig:venn}
\end{figure}

\begin{table}[t]
    \centering
    \caption{Comparison of the number of unique type error fixes and template coverage between \tool and PyTER.}
    \scalebox{0.9}{
    \begin{tabular}{ccccc}
    \toprule
        \multirow{2}*{\textbf{Approach}} & \multicolumn{2}{c}{\textbf{\textsc{TypeBugs}}} & \multicolumn{2}{c}{\textbf{\textsc{BugsInPy}}} \\
        & \#Unique & Coverage  & \#Unique & Coverage \\
    \midrule
        \tool & 24 & 83 (76.1\%) & 16 & 40 (74.1\%)  \\
        PyTER & 10 & 46 (42.2\%) & 5 & 18 (33.3\%) \\
        \bottomrule
    \end{tabular}}
    \label{tab:coverage}
\end{table}

\textbf{Comparison with Learning-based Approaches.} From Table~\ref{tab:main_res} we can see that \tool, Codex and AlphaRepair generally perform much better than CoCoNuT, indicating the superior performance of prompt-based approaches. When comparing \tool with AlphaRepair which adopts general domain-unaware prompt templates, we find that \tool achieves a $1\times \sim 4\times$ larger fix rate than AlphaRepair. This indicates that general domain-unaware prompt templates such as randomly replacing several tokens in code can hardly handle complicated type errors. Compared with the most advanced code pre-trained model Codex, \tool still obtains a significant improvement by fixing 16 and 9 more type errors than Codex in \textsc{TypeBugs} and \textsc{BugsInPy}, respectively. This improvement further demonstrates the importance of domain knowledge for repairing type errors, even though Codex has a much larger parameter size (12B) than that (220M) of the CodeT5 model utilized by \tool.

In addition to the total number of type errors fixed by each approach, we further evaluate the number of unique type error fixes. Fig.~\ref{fig:venn} presents the unique type errors that \tool and three learning-based approaches can correctly fix in the format of Venn diagrams. We observe that \tool obtains 16 and 10 unique type error fixes in \textsc{TypeBugs} and \textsc{BugsInPy}, respectively, while other approaches only obtain $0 \sim 3$ unique bug fixes in two benchmarks. This indicates that the contribution of domain-aware fix templates cannot be replaced by the combination of existing learning-based approaches.

\begin{table}[t]
    \centering
    \caption{Statistics of fix templates mining in \tool.}
    \scalebox{0.9}{
    \begin{tabular}{cccc}
    \toprule
    \multirow{2}*{\textbf{Category}} & \multirow{2}*{\textbf{\#Instances}} & \multirow{2}*{\tabincell{c}{\textbf{\#Clustering Trees} \\ \textbf{(>1/>5)}}} & \multirow{2}*{\textbf{Mining Time/s}}\\
    & & & \\
    \midrule
       Add  & 2,656  & 150/27 & 2,819\\
       %\midrule
     Remove & 570 & 10/5 & 59\\
     %\midrule
     Insert & 1,648 & 350/47 & 659\\
     %\midrule
     Replace & 6,107 & 184/70 & 32,621\\
     \bottomrule
    \end{tabular}}
    
    \label{tab:mining}
\end{table}

\begin{table}[t]
    \centering
    \caption{Ablation results.}
    \scalebox{0.9}{
    \begin{tabular}{ccccc}
    \toprule
        & \multicolumn{2}{c}{\textbf{\textsc{TypeBugs}}} & \multicolumn{2}{c}{\textbf{\textsc{BugsInPy}}} \\
         &  \textbf{\#Correct} & \textbf{\#Plausible} & \textbf{\#Correct} & \textbf{\#Plausible} \\
    \midrule
       No Template  & 19  & 41 & 6 & 20 \\
    \midrule
       Add  & +11 & +10 & +3 & +5 \\
       Remove  & +1 & +1 & +0 & +0 \\
       Replace  & +6  & +8 & +4 & +9 \\
       Insert & +18  & +16 & +13 & +17 \\
    \midrule
        \textbf{Total} & 55  & 76 & 26 & 41 \\
    \bottomrule
    \end{tabular}}
    \label{tab:ablation}
\end{table}

\answer{1}{\tool successfully fixes 55 and 26 bugs in two benchmarks, outperforming state-of-the-art approaches by at least 14 bugs and 9 bugs, respectively. Meanwhile, \tool obtains the most unique type error fixes in two benchmarks.}

\subsection{RQ2: Capability of \tool to Mine Fix Templates}

To comprehensively investigate the capability of \tool to mine fix templates, we focus on the performance of \tool in template mining and the usefulness of fix templates mined by \tool.

Table~\ref{tab:mining} presents the performance of \tool in fix template mining process. Starting with thousands of existing bug fixes, \tool can mine $10 \sim 350$ clustering trees. After discarding the clustering trees with occurrence frequency lower than a threshold (5 in this paper), \tool finally gets $5 \sim 70$ clustering trees. The mining process generally takes shorter than one minute to at most nine hours. Table~\ref{tab:coverage} presents the template coverage achieved by \tool and PyTER. From it we can observe that fix templates mined by \tool can cover about 75\% of type errors in two benchmarks while the manually defined fix templates in PyTER can only cover about $30\% \sim 40\%$ of type errors.

To further study how fix templates mined by \tool can help the patch generation process, we conduct an ablation study on fix templates under each category. Following previous study~\cite{xia22less}, we start with the case that no fix template is applied, i.e., the CodeT5 model is asked to generate a complete new line to replace the original buggy line. We then gradually apply fix templates under \textit{Add}, \textit{Remove}, \textit{Replace} and \textit{Insert} categories, and observe the number of new correct and plausible patches, respectively. We show the results in Table~\ref{tab:ablation}. We can find that all four categories of fix templates contribute to generating correct patches. This demonstrates the contribution of domain knowledge stored in the fix templates. We also note that fix templates under \textit{Insert} and \textit{Add} categories contribute the most. The reason could be attributed to that developers often add guards to guarantee the desired types or directly convert input types into desired types when fixing type errors.

\answer{2}{\tool achieves a template coverage of about 75\% on both benchmarks. Ablation results also demonstrate the usefulness of fix templates mined by \tool under each category.}

\subsection{RQ3: Limitations of \tool}
Our experiments also show the limitations of \tool as it cannot fix all type errors in two benchmarks. By analyzing the type errors that \tool cannot fix in two benchmarks, we conclude two possible limitations.

The first limitation is that \tool cannot always find matched fix templates to the current buggy program. Based on Table~\ref{tab:coverage}, we can find that even if fix templates mined by \tool can cover as many as 75\% cases in two benchmarks, there exist a few cases ($\sim$25\%) that do not share similar patterns with instances in the training set. An example is illustrated in the first type error of Listing~\ref{lst:ex2}. To fix this type error, the developer changes a list comprehension into an attribute access, which does not appear in the training set. \tool thus cannot find proper fix templates for this type error and fails to fix it. This limitation can be mitigated by adapting \tool to new datasets, so that \tool can mine new fix templates to improve the template coverage.

\begin{lstlisting}[label=lst:ex2, caption=Two type errors that \tool fail to fix]
#Type Error 1: apache/airflow:892d4d
if conf.getboolean('core', 'store_dag_code',\
fallback=False):
<@\colorbox{red!30!}{- \quad \quad DagCode.bulk\_sync\_to\_db([dag.fileloc for dag in orm\_dag])}@>
<@\colorbox{green!30!}{+ \quad \quad DagCode.bulk\_sync\_to\_db([orm\_dag.fileloc])}@>
#Type Error 2: pandas-dev/pandas:a3e903
elif (is_extension_array_dtype(left) or\
<@\colorbox{red!30!}{- is\_extension\_array\_dtype(right)):}@>
<@\colorbox{green!30!}{+ (is\_extension\_array\_dtype(right) and not is\_scalar(right))):}@>
       return dispatch_to_extension_op(op, left, right)
\end{lstlisting}

The second reason is that the CodeT5 model \tool uses sometimes cannot generate the correct patches even if the correct fix template is given. By comparing Table~\ref{tab:main_res} and Table~\ref{tab:coverage}, we can find that fix templates mined by \tool can cover 83 bugs in \textsc{TypeBugs} but only 55 of them are correctly fixed. This indicates the limitations of CodeT5 when generating candidate patches from code prompts. We also show an example as the second type error of Listing~\ref{lst:ex2}. In this type error, we need to add a new condition as the guards and this fix pattern is commonly used in the wild. However, CodeT5 cannot give \textit{is\_scalar} as the new condition and thus \tool fails to fix this type error. We believe this limitation can be mitigated by using more advanced code pre-trained models, as the parameter size of CodeT5 is only 220M.
% the CodeT5 model \tool utilizes in this paper has a parameter size of only 220M.

\answer{3}{\tool sometimes fails to fix type errors due to the limited performance of pre-trained code models and a few cases ($\sim$ 25\%) that mined fix templates cannot cover.}

%\section{Discussion} \label{sec:discussion}
\section{Related Work}\label{sec:literature}

\subsection{Automatic Program Repair}
As an important method to improve the reliability of software, automatic program repair (APR) has draw a lot of attention~\cite{gazzola19automatic, monperrus18automatic}  in recent years. Currently, most APR approaches can be classified into rule-based approaches and learning-based approaches.

\textbf{Rule-based Approaches.}
Rule-based APR approaches leverage pre-defined templates and rules to generate patches for bugs via static and dynamic analysis. There are a series of rule-based APR approaches designed for Java programs via constant-solving~\cite{xuan17nopol, mechtaev16angelix}, fuzzing~\cite{gao19crash}, testcase generation~\cite{xiong17precise}, bytecode mutation~\cite{ghanbari19practical}, and for the memory bugs of C programs~\cite{gao15safe, hong20saver, hu19re, lee18memfix, yan16automated}. For Python programs, PyTER~\cite{pyter} utilizes nine pre-defined templates with type-aware fault localization to repair type errors. Traditional rule-based approaches are domain-aware but can be used to fix a limited amount of real-world bugs. \tool addresses this challenge by automatically mining fix templates from real-world bug fixes. Different from previous work on mining code edit patterns~\cite{getafix, sakkas20type, rolim17learning, rolim21learning}, \tool implements a novel fix template design to handle type errors at different levels.

\textbf{Learning-based Approaches.}
Learning-based APR approaches become quite popular and demonstrate their superior performance recently. Motivated by the study~\cite{Wang2020RethinkingTV, Wang2022UnderstandingAI, yang20a} of neural machine translation (NMT)~\cite{sutskever14sequence}, there are many research efforts being devoted to NMT-based APR approaches. SequenceR~\cite{chen21sequencer} presents a sequence-to-sequence LSTM model for program repair. DLFix~\cite{li20dlfix} leverages tree-based RNN to transform code inputs and generate patches. CoCoNuT~\cite{Lutellier20coconut} separates the context and buggy line in NMT-based APR. CURE~\cite{jiang21cure} is the first approach that integrates pre-trained models in NMT-based APR, followed by work~\cite{drain21deepdebug, mashhadi21applying}. Recoder~\cite{zhu21a} generates code edits instead of modified code. RewardRepair~\cite{ye22neural} uses execution-based backpropagation to improve the compilation rate of patches generated by NMT-based APR approaches. AlphaRepair~\cite{xia22less} is the first prompt-based APR approach that transforms the APR problem into a fill-in-the-blank problem. However, general domain-unaware prompts AlphaRepair uses can hardly handle complicated type errors. \tool addresses the problem by incorporating domain knowledge with proposed fix templates and mining them through existing type error fixes.

\subsection{Pre-trained Language Models}
Making use of large-scale unlabeled data in the wild, pre-trained language models are proven to be quite effective in the natural language processing (NLP) field. Inspired by such a success, many code pre-trained models are proposed. Most code pre-trained models utilize the same Transformer~\cite{vaswani17attention} structure. CodeBERT~\cite{feng20codebert} is a BERT-style model pre-trained on both programming languages and natural languages. GraphCodeBERT~\cite{guo21graphcodebert} utilizes the data flow graphs in the pre-training stage. CodeGPT~\cite{codegpt} is a GPT~\cite{gpt}-style model pre-trained on Python and Java functions. Codex~\cite{codex} is a GPT-style model created by fine-tuning the GPT3~\cite{gpt3} model to generate Python functions. CodeT5~\cite{wang21codet5} is an encoder-decoder model pre-trained on CodeSearchNet~\cite{codesearchnet}. Recently prompt learning~\cite{liu21pre, wang22no} draws a lot of attention. By transforming downstream tasks into fill-in-the-blank problems, code pre-trained models can be directly used for automatic program repair via predicting appropriate code tokens required to fix bugs~\cite{xia22less, prenner22can, sobania23an}.

\section{Conclusion}
We propose a domain-aware prompt-based approach named \tool for repairing Python type errors. \tool improves prompt-based approach by incorporating domain-aware fix templates. \tool implements a novel fix template design to handle type errors at different levels, and mines fix templates via a novel hierarchical clustering algorithm. \tool incorporates domain knowledge into code prompts by applying fix templates into buggy code and invokes code pre-trained models to generate candidate patches from code prompts. Experiments demonstrate the effectiveness of \tool and the usefulness of fix templates mined by \tool.

\section{Acknowledgments}
The authors would like to thank the efforts made by anonymous reviewers. The work described in this paper was supported by the Research Grants Council of the Hong Kong Special Administrative Region, China (No. CUHK 14206921 of the General Research Fund). The work was also supported by National Natural Science Foundation of China under project (No. 62002084), Natural Science Foundation of Guangdong Province (Project No. 2023A1515011959), Shenzhen Basic Research (General Project No. JCYJ20220531095214031), and Key Program of Fundamental Research from Shenzhen Science and Technology Innovation Commission (Project No. JCYJ20200109113403826). Any opinions, findings, and conclusions or recommendations expressed in this publication are those of the authors, and do not necessarily reflect the views of the above sponsoring entities.

\section{Data Availability}
The implementation of \tool and experiment results discussed in this paper are available at https://github.com/JohnnyPeng18/TypeFix.

%\tool mines domain-aware fix templates via a novel hierarchical clustering algorithm and applies them to generate code prompts for the prediction of code pre-trained models to generate patch candidates. %We evaluate the performance of \tool on two benchmarks \textsc{TypeBugs} and \textsc{BugsInPy}. \tool outperforms state-of-the-art rule-based and learning-based approaches by at least 14 and 9 bugs, respectively. \tool achieves a template coverage of 75\%, which is $1.8\times \sim 2.2\times$ of PyTER. The ablation study and case study both demonstrate the usefulness of fix templates mined by \tool.

\newpage
\bibliographystyle{ACM-Reference-Format}
\bibliography{ref}

\end{document}